\documentclass[preprint,12pt]{aastex}
\usepackage{psfig}
\newcommand{\be}{\begin{equation}}
\newcommand{\ee}{\end{equation}}
\newcommand{\bea}{\begin{eqnarray}}
\newcommand{\eea}{\end{eqnarray}}
\newcommand{\rmscr}[1]{{\hbox{\scriptsize \rm{#1}}}}

\bibpunct[,]{(}{)}{;}{a}{}{,}
\begin{document}

\title{Constraining Alternate Models of Black Holes: Type I X-ray
Bursts on Accreting Fermion-Fermion and Boson-Fermion Stars}

\author{Ye-Fei Yuan}

\affil{Center for Astrophysics, Univ. of Science and Technology,
Hefei, Anhui 230026, P. R. China; yfyuan@ustc.edu.cn}

\author{Ramesh Narayan}

\affil{Harvard-Smithsonian Center for Astrophysics, 60 Garden Street,
Cambridge, MA 02138, U.S.A.; narayan@cfa.harvard.edu}

\author{Martin J. Rees}

\affil{Institute of Astronomy, Univ. of Cambridge, Madingley Road,
Cambridge CB3 0HA, England; mjr@ast.cam.ac.uk}

\begin{abstract}

The existence of black holes remains open to doubt until other
conceivable options are excluded.  With this motivation, we consider a
model of a compact star in which most of the mass consists of dark
particles of some kind, and a small fraction of the mass is in the
form of ordinary nucleonic gas.  The gas does not interact with the
dark matter other than via gravity, but collects at the center as a
separate fermionic fluid component.  Depending on whether the dark
mass is made of fermions or bosons, the objects may be called
fermion-fermion stars or boson-fermion stars, respectively.  For
appropriate choices of the mass of the dark matter particles, these
objects are viable models of black hole candidates in X-ray binaries.
We consider models with a dark mass of $10M_\odot$ and a range of gas
mass from $10^{-6}M_\odot$ to nearly $1M_\odot$, and analyse the
bursting properties of the models when they accrete gas.  We show that
all the models would experience thermonuclear Type I X-ray bursts at
appropriate mass accretion rates.  Since no Type I bursts have been
reported from black hole candidates, the models are ruled out.  The
case for identifying black hole candidates in X-ray binaries as true
black holes is thus strengthened.

\end{abstract}


\section{Introduction}

Eighteen excellent black hole candidates have been discovered so far
in X-ray binaries (XRBs, McClintock \& Remillard 2003).  The compact
stars in these binaries have masses that exceed the maximum mass of a
neutron star (Glendenning 2000); therefore, it is generally assumed
that they must be black holes.  The mass measurements certainly
provide a strong argument for considering the objects to be black
holes.  However, it would be prudent to keep an open mind on this
matter and to consider the possibility that the objects may be some
kind of exotic stars that are composed of an as-yet unidentified form
of exotic material.  Until such a model is ruled out, the objects must
be treated only as black hole candidates (BHCs), not true black holes.

Glendenning (2000) has discussed a number of forms of exotic matter
that might be present in compact stars, and describes models made up
of these kinds of matter.  All of the models have a maximum mass that
is well under $3M_\odot$.  The models are, therefore, not relevant for
BHCs in XRBs.  There are other forms of compact stars, however, e.g.
Q-stars (Miller, Shahbaz \& Nolan 1998, and references therein), that
are in principle consistent with the observed objects.  Can one show
that such objects are ruled out?  One way to do this is to demonstrate
that BHCs do not have hard surfaces.  If we could show this, then the
objects must have event horizons and must, therefore, be black holes.
Beginning with the work of Narayan, Garcia \& McClintock (1997, 2002)
and Garcia et al. (2001), a number of studies have attempted to prove
this result (e.g., Sunyaev \& Revnivtsev 2000; Narayan \& Heyl 2002;
Done \& Gierlinsky 2003).  We are concerned in this paper with the
work of Narayan \& Heyl (2002), who argued that BHCs should exhibit
Type I X-ray bursts if they have surfaces.  We give an update on that
work in \S4 of this paper and show that the calculations rule out
models of BHCs in which the objects have a hard surface on which
accreting gas can collect.

This still leaves open the possibility that BHCs may be made up of
some kind of exotic dark matter with which normal gas does not
interact.  That is, the dark matter may be ``porous'' and permit
accreting gas to fall through and to collect at the center.  The dark
matter and the fermionic gas would then behave as two independent
fluids that interact only via gravity.  Such models have been
discussed in the literature, both for fermionic dark matter and
bosonic dark matter (Lee \& Pang 1987; Zhang 1988; Henriques et
al. 1989, 1990a,b; Jin \& Zhang 1989, 1999).  We refer to these
objects as fermion-fermion stars and boson-fermion stars,
respectively, where the first half of the name refers to the nature of
the dark matter, and the second ``fermion'' in each name corresponds
to the nucleonic gas component.  If BHCs consist of either
fermion-fermion or boson-fermion stars, would they have Type I bursts
when they accrete gas from a normal binary companion star?  We attempt
to answer this question.

We describe in \S\S2,3 models of fermion-fermion and boson-fermion
stars consisting of $10M_\odot$ of dark matter and various amounts of
gas mass $M^g$.  We calculate how relevant properties of the gas
sphere, such as the radius, surface gravity, redshift, etc., vary as a
function of $M^g$.  Then, in \S4, we describe the bursting properties
of these models.  We conclude in \S5 with a discussion of the results.
Appendix A relates the accretion efficiency of a fermion-fermion star
to the redshift at the surface of the gas sphere.

\section{Fermion-Fermion Stars}

We consider a spherically-symmetric fermion-fermion star consisting of
dark non-interacting fermions plus ordinary gas.  We assume that the
fermion particles have a mass $m_f$ and that the star has a total of
$N^f$ fermions, corresponding to a dark ``baryonic mass'' of
$M^f=N^fm_f$.  The ordinary gas component has a baryonic mass
$M^g=N^gm_n$, where $m_n=936$ MeV is the average mass of a nucleon,
assuming the initial accreted gas contains 70\% Hydrogen and 30\%
Helium and other elements (Narayan \& Heyl 2003).  The gravitational
mass of the combined star is $M_{\rm grav}$.

We solve for the equilibrium structure of the fermion-fermion star
using the field equations of General Relativity.  In Schwarzschild
coordinates, the line element of a spherically symmetric,
time-independent spacetime can be written as
\begin{equation}
ds^2=-B(r)dt^2+A(r)dr^2+r^2d\Omega^2 ,
        \label{eq:metric}
\end{equation}
where $A(r)\equiv 1/(1-2GM(r)/r)$, $G$ is the gravitational constant,
and $M(r)$ is the gravitational mass enclosed within radius $r$. For
an asymptotically flat spacetime, we have $A(\infty)\rightarrow 1$ and
$B(\infty)\rightarrow 1$.  For the specific case of the well-known
Schwarzschild metric, $B(r)=1/A(r)=(1-2GM_{\rm grav}/r)$, where $M_{\rm grav}$
is the total gravitational mass.  Since our fermion-fermion star has a
spatially distributed mass, we employ the more general metric written
in equation (\ref{eq:metric}).

The metric functions $A(r)$ and $B(r)$ are determined by the Einstein
field equations,
\begin{equation}
R_{\mu\nu}-\frac{1}{2}g_{\mu\nu}R=8\pi G[T^f_{\mu\nu}+T^g_{\mu\nu}],
\end{equation}
where $T^f_{\mu\nu}$ is the energy-momentum tensor of the fermions and
$T^g_{\mu\nu}$ is that of the normal nuclear matter.  We treat both
the dark fermions and the gas as perfect fluids,
\begin{equation}
T^i_{\mu\nu}(\rho^i,p^i)=(\rho^i+p^i)U_{\mu}U_{\nu}+g_{\mu\nu}p^i,
~~~(i=f,g),
        \label{eq:einstein}
\end{equation}
where $U_{\mu}$ is the four-velocity of the fluid.  From equations
(\ref{eq:metric}) and (\ref{eq:einstein}), we have the following
structure equations, which are an obvious generalization of the
Tolman-Oppenheimer-Volkoff equations for the case of a single fluid,
\begin{eqnarray}
A^{'}(r)  &=& 8\pi G rA^2(r)[\rho^f(r)+\rho^g(r)]-\frac{A(r)}{r}[A(r)-1], \\
B^{'}(r)  &=& 8\pi G rA(r)B(r)[p^f(r)+p^g(r)]+\frac{B(r)}{r}[A(r)-1], \\
p^{f'}(r) &=& -\frac{1}{2} \frac{B^{'}(r)}{B(r)}[\rho^f(r)+p^f(r)], 
        \label{eq:dpdr} \\
p^{g'}(r) &=& -\frac{1}{2} \frac{B^{'}(r)}{B(r)}[\rho^g(r)+p^g(r)], \\
N^{f'}(r) &=& 4 \pi r^2 \sqrt{A(r)} n^f(r), \\
N^{g'}(r) &=& 4 \pi r^2 \sqrt{A(r)} n^g(r),
\end{eqnarray}
where primes denote derivatives with respect to radius $r$,
$\rho^f(r)$ and $\rho^g(r)$ are the proper mass densities of the
fermions and the gas, $p^f(r)$ and $p^g(r)$ are the corresponding
pressures, and $N^f(r)$ and $N^g(r)$ are the total numbers of fermions
and gas nucleons enclosed within radius $r$.  

The boundary conditions are as follows,
\begin{equation}
A(0)=1,~B(0)=b_0,~p^f(0)=p^f_c,~p^g(0)=p^g_c,
~N^f(0)=0,~N^g(0)=0,
\end{equation}
where $p^f_c$ and $p^g_c$ are the central pressures of the fermion
fluid and the gas, respectively.  Because the equations are linear in
$B(r)$, $b_0$ can be assigned an arbitrary value and later rescaled so
as to satisfy the asymptotic flatness condition, $B(\infty)=1$.  The
two central pressures are free parameters that are determined by
applying boundary conditions at large $r$ on the total baryonic masses
of the two fluids.

In order to solve the
above coupled equations, we need to know the equations of state of the
fermions and the gas,
\begin{eqnarray}
\rho^i=\rho^i(n^i), \\
p^i=p^i(n^i),
\end{eqnarray}
where $n^i$ is the number density of particles of type $i$ ($=f, g$).
For simplicity, we treat the dark fermions as a non-interacting gas
whose equation of state (EOS) may be written in parametric form (see
Chandrasekhar 1935; Oppenheimer \& Volkoff 1939),
\begin{eqnarray}
\rho &=& \frac{m_f^4}{32 \pi^2}(\sinh t-t), \\
p &=& \frac{1}{3} \frac{m_f^4}{32 \pi^2}(\sinh t-8\sinh(t/2)+3t),
\end{eqnarray}
where the parameter $t$ is defined by
\begin{equation}
t=4\log \left\{ \frac{p_F}{m_f} + \left[ 1+
        \left(\frac{p_F}{m_f}\right)^2 \right]^{1/2} \right\},
\end{equation}
and $p_F$ is the Fermi momentum. The number density of fermions is
\begin{equation}
n^f=\frac{p_F^3}{3\pi^2}.
\end{equation}
For the normal gas component, we use the FPS equation of state (EOS,
Lorenz, Ravenhall \& Pethick 1993) when the density is below the
nuclear density $\rho_{\rm nuc}$, matching it smoothly to the
relativistic mean field theory (RMFT) EOS at higher density, taking
the so-called GM1 set of the coupling constants (Glendenning \&
Moszkowski 1991).  In practice, the maximum central density of the gas
in the models is below $2\rho_{\rmscr{nuc}}$, so different sets of the
coupling constants give almost the same results.

Starting from $r=0$, the equations for the fermion pressure and
fermion number are integrated out until the pressure goes to zero.
The radius at which this happens is identified as the surface of the
fermion star.  So too for the gas.  In calculating the fermion-fermion
models, we first assumed that we have a pure dark fermion star with no
gas.  We adjusted the central pressure $p^f(0)$ of this model such
that the asymptotic gravitational mass $M_{\rm grav}^f(r\to\infty)$ is
equal to $10M_\odot$.  We selected the mass of the fermion particles
to be 223 MeV.  For this choice of mass, the maximum mass of a pure
fermion star is $12.61M_\odot$.  A pure fermion star with a mass of
$10M_\odot$ made of these particles is fairly compact: $R_f=252$ km,
$R_f/R_S = 8.56$ in Schwarzschild units.  Therefore, the model is
interesting as a possible description of a black hole candidate in an
X-ray binary.  Figure 1 shows the metric quantities $A(r)$ and $B(r)$
as functions of radius for this model.

We then added different amounts of gas to the above model, from
baryonic mass $M^g=10^{-6}M_\odot$ upto the maximum allowed mass
beyond which the object becomes a black hole, and calculated the
structure of the combined fermion-fermion star.  In each model, we
adjusted the central pressures $p^f$ and $p^g$ such that the total
number of dark fermions is equal to the same $N^f$ obtained for the
pure fermion model, and the baryonic mass of the gas is equal to the
desired value.  Figure 1 shows the results for the specific case when
the baryonic mass of the gas is equal to $0.7M_\odot$.  Note that the
gas sphere is very much more compact than the fermion sphere in which
it is embedded.  This is the case for all the models we have
calculated.

For each choice of the baryonic mass of the gas $M^g$, we obtained the
radius $R_g$ of the gas component (the radius at which the gas
pressure vanishes), the surface gravitational acceleration $g$, and
the redshift at the gas surface $z$.  The redshift $z(r)$ at radius
$r$ is defined as
\begin{equation}
z(r) \equiv \frac{1}{\sqrt{B(r)}}-1.
\end{equation}
For $g$, we rewrite equations (\ref{eq:dpdr}) and (\ref{eq:dpdx}) in
the following form,
\begin{equation}
\frac{dp^g}{d\Sigma}=-\frac{1}{2}\frac{B^{'}(r)}{B(r)}
        \frac{1}{\sqrt{A(r)}} \left(1+\frac{p^g}{\rho^g} \right)
        \equiv g,
        \label{eq:geff}
\end{equation}
which defines the effective gravitational acceleration. At the surface
of the accreted gas, particles are nonrelativistic, thus $p^g/\rho^g
\ll 1$.  The quantity $\Sigma$ in equation (\ref{eq:geff}) is the
surface mass density which is defined by
$d\Sigma(r)=\rho^g(r)\sqrt{A(r)}dr$.  We also calculated an effective
accretion efficiency $\eta\equiv L_{\rm acc}/\dot Mc^2$, which
measures the fraction of the rest mass energy of accreting matter that
is released when gas accretes on the fermion-fermion star.  This is
obtained in terms of the gravitational mass of the star as
follows:
\begin{equation}
\eta = 1-{d M_{\rm grav} \over dM^g}.
\end{equation}
We show in Appendix A that $\eta$ is given very simply in terms of the
redshift at the surface of the gas sphere.  The relation we derive is
a relativistic generalization for a two-fluid system of a Newtonian
result obtained by Rosenbluth et al. (1973).

Figure 2 shows the results of the calculations.  The four panels show
the variations of the radius of the gas sphere, the gravitational
acceleration at its surface, the redshift, and the accretion
efficiency, as functions of the gas baryonic mass.  With increasing
gas mass, the radius of the gas sphere decreases, the gravitational
acceleration increases, and the redshift and accretion efficiency
increase.  The wiggle in the curves at small masses corresponds to a
switch between two solution branches.  Note that the maximum mass in
this sequence of models is $0.782M_\odot$; beyond this mass, the
object collapses to a black hole.

\section{Boson-Fermion Stars}

Boson stars are macroscopic quantum states resulting from the
self-gravitational equilibrium of boson fields (see the reviews by
Jetzer 1992; Lee 1992; Liddle \& Madsen 1992; Mielke \& Schunke 1998;
Schunke \& Mielke 2003).  There is no concept of an equation of state
for these systems, as they are pure quantum systems that are held up
against gravitational collapse by the Heisenberg uncertainty
principle. The pioneering studies in this field were done by Kaup
(1968) and Ruffini \& Bonazzola (1969). The main findings in these two
seminal works are that (i) the mass of a boson star is of the order of
$M^2_{\rmscr{Pl}}/m_b$, and (ii) its characteristic size is of the
order of the de Broglie wavelength of bosons $1/m_b$, where $m_b$ is
the mass of the boson particle and $M_{\rmscr{Pl}}$ is the Planck
mass.  Colpi, Shapiro \& Wasserman (1986) introduced the idea of
self-interaction of the scalar particles and found that
self-interacting boson stars have masses of the order of
$\Lambda^{1/2}M^2_{\rmscr{Pl}}/m_b$, where $\Lambda$ is a
dimensionless quantity which characterizes the strength of the self
interactions.  For $\Lambda^{1/2}\gg1$, this scaling breaks down, and
the mass is instead $\sim M^3_{\rmscr{Pl}}/m^2$, which is similar to
the result for fermion stars.  

Boson stars have several interesting/unique characteristics, including
their transparency to photons and baryonic matter, their ability to be
non-singular even for masses larger than the maximum mass of neutron
stars (hence the interest in these objects as a model of BHCs), and
the presence of a different metric than the Schwarzschild metric.
Various authors have discussed observational consequences of boson
star, such as the \v{C}erenkov effect, gravitational lensing, the
rotation curves of accreted matter, and the gravitational redshift of
the radiation emitted within the effective radius of boson stars
(Schunk \& Mielke 2003).  Also, the possibility of supermassive
nonbaryonic stars (boson stars or neutrino balls) insteading of
supermassive black holes existing in the nuclei of galaxies has
attracted attention (Schunk \& Liddle 1997, 1998; Torres, Capozziello
\& Lambiase 2000; Tsiklauri \& Viollier 1998).  It has been argued
that the line profile of an emission line from an accretion disk
around a supermassive boson star may have signatures that might help
to identify the existence of boson stars (Lu \& Torres 2003).

The observational signatures of boson stars in X-ray binaries with
BHCs has not been discussed very much in the literature.  We show in
this paper that such stars will produce Type I X-ray bursts with clear
signatures.

Henriques (1989, 1990a,b) considered the possibility of compact stars
that contain both bosons and fermions and wrote down the structural
equations for such objects.  We borrow from their analysis in what
follows.  We analyse a boson-fermion star that consists of dark bosons
plus ordinary fermionic gas.  As in the case of the fermion-fermion
star, we write the number of bosons as $N^b$ and the baryonic mass of
the bosons as $M^b=N^bm_b$.

The energy-momentum tensor for bosons is completely different from
that of fermions.  For a massive self-interacting scalar field, the
Lagrangian reads (Colpi et al. 1986)
\begin{equation}
L = -\frac{1}{2} g^{\mu \nu} D_{\mu}
\phi D_{\nu} \phi -\frac{1}{2} m_b^2 \phi^2-\frac{1}{4} \lambda \phi^4,
\end{equation}
where $\lambda$ is the self-interaction coupling constant. According
to Noether's theorem, the energy-momentum tensor can be written as
\begin{equation}
T^b_{\mu\nu}(\phi)=D_{\mu}\phi D_{\nu}\phi -\frac{1}{2}g_{\mu \nu}
(g^{\rho \sigma}D_{\rho}\phi D_{\sigma} \phi+m_b^2\phi^2
+\frac{1}{2}\lambda\phi^4).
\end{equation}
In order to solve the Einstein equations, it is reasonable to assume
that
\begin{equation}
\phi(r,t)=\Phi(r)e^{-i\omega t},
\end{equation}
where $\Phi(r)$ is a real function.  The structure equations can then
be obtained from the Einstein equations (Henriques 1989),
\begin{eqnarray}
A^{'}(x)&=&xA^2(x)\left[2\bar{\rho}^g(x)+\left(\frac{\Omega^2}{B(x)}+1\right)\sigma^2
       +\frac{\Lambda}{2}\sigma^4+\frac{\sigma^{'}}{A(x)}\right]
       -\frac{A(x)}{x}[A(x)-1], \\
       B^{'}(x)&=&xA(x)B(x)\left[2\bar{p}^g(x)+\left(\frac{\Omega^2}{B(x)}-1\right)\sigma^2
       -\frac{\Lambda}{2}\sigma^4+\frac{\sigma^{'}}{A(x)}\right]
       +\frac{B(x)}{x}[A(x)-1], \\ \sigma^{''}&=&-\left( \frac{2}{x}
       +\frac{B^{'}(x)}{2B(x)} -\frac{A^{'}(x)}{2A(x)} \right)
       \sigma^{'} -A\left[ \left(\frac{\Omega^2}{B(x)}-1 \right)
       \sigma -\Lambda \sigma^{3} \right], \\
       \bar{p}^{g'}(x)&=&-\frac{1}{2}
       \frac{B^{'}(x)}{B(x)}[\bar{\rho}^g(x)+\bar{p}^g(x)],
       \label{eq:dpdx} \\ N^{b'}(x) &=& \frac{1}{G} x^2
       \frac{\Omega}{\sqrt{B(r)}} \sqrt{A(x)} \sigma^2, \\ N^{g'}(x)
       &=& 4 \pi x^2 \sqrt{A(x)} \bar{n}^g(x),
\end{eqnarray}
where all physical quantities have been redefined in dimensionless
form,
\begin{eqnarray}
x&=&m_br, ~~\sigma=\sqrt{4\pi G}\Phi,
~~\Omega=\omega/m_b,~~\Lambda=\lambda/(4\pi
Gm_b^2), \nonumber \\
\bar{\rho}&=&4\pi G\rho /m_b^2, ~~\bar{p}=4\pi G p/m_b^2, ~~\bar{n}^g=n^g
m_b^{-3}.
        \label{eq:dimless}
\end{eqnarray}

The boundary conditions are as follows,
\begin{eqnarray}
A(0) &=& 1, ~~B(0)=b_0, ~~\sigma(0)=\sigma_0, ~~\sigma^{'}(0)=0,\nonumber \\
p^g(0) &=& p^g_c, ~~N^b(0)=0, ~~N^g(0)=0, ~~\Omega^2(0)/b_0=c_0 .
\end{eqnarray}
In addition, the total gravitational mass $M_{\rm grav}$ is given by
the asymptotic value of $A$ at large $r$: $A(\infty) \to (1-GM_{\rm
grav}/r)^{-1}$.  As in the case of the fermion-fermion star, the
equations are linear in $b_0$.  Hence, this parameter may be given any
value and normalized after the calculations have been done so as to
achieve $B(\infty)=1$.  Because the differential equation in $\sigma$
is second-order, we require two boundary conditions, one on $\sigma$
and another on $\sigma^{'}$.  As in the case of the fermion star, two
parameters, namely $\sigma_0$ and $p_c^g$, are determined by requiring
the total boson and gas masses to equal specified values.  In
addition, there is an extra parameter $c_0$ which behaves like an
eigenvalue.  This parameter is adjusted so as to obtain a
wave-function $\sigma$ that is both nodeless and behaves regularly at
infinity, i.e., $\sigma$ should remain positive for all $x$ and should
asymptote smoothly to 0 as $x\to\infty$.  (If $c_0$ has the wrong
value, $\sigma$ either crosses 0 and goes negative at a finite radius,
or diverges as $x\to\infty$.)

As in the case of the hybrid fermion star, we first solve for a pure
boson star that has a gravitational mass of $10M_\odot$.  We selected
a boson particle mass of $m_b=2.4\times10^{-17}$ MeV and a coupling
constant of $\Lambda=100$.  For these parameters, the maximum mass of
a pure boson star is $12.57M_\odot$.  Figure 1 shows the variations of
$A(R)$ and $B(R)$ with $R$ for the pure boson star.  The object does
not have a precise surface since the wave-function extends to
infinity.  If we define the radius as the value of $r$ at which 99.9\%
of the mass is enclosed, then the radius is 153.3 km.

Having solved for the structure of the pure boson star, we then added
various amounts of gas from $10^{-6}M_\odot$ up to the maximum allowed
mass of $0.863M_\odot$ (above which the object becomes a black hole)
and solved for the structure of the combined boson-fermion star.
Figure 1 shows the solution for the metric functions when
$M^g=0.7M_\odot$.  Note that, even though the boson component does not
have a precise surface, the gas component behaves like a fermion
object and does have a surface at which the pressure goes to zero.
The radius of this surface, the surface gravitational acceleration,
the redshift and the accretion efficiency are shown as functions of
$M^g$ in Fig. 3.

\section{Type I X-ray Bursts in Black Hole Candidates}

If black hole candidates (BHCs) in X-ray binaries are not true black
holes with event horizons, then they must have surfaces of some kind.
As discussed in \S1, we may consider two general classes of models.
In one class of models, BHCs are made of exotic matter and have a hard
surface at some radius $R_{\rm BHC}$.  When gas accretes on such a
star, it is arrested at the surface and piles up as a layer of normal
matter.  Narayan \& Heyl (2002) showed that the accreting layer will
generally produce Type I X-ray bursts.

Type I bursts in accreting neutron stars were originally discovered by
Grindlay et al. (1976) and have been studied intensively for many
years (see Lewin, van Paradijs \& Taam 1993 for a review of the
observations, and Bildsten 1998 for a discussion of the physics).
When gas accretes and accumulates on a compact star and is compressed
by the strong gravity, thermonuclear reactions are ignited in the gas.
In many cases (but not all, see Narayan 2003) the nuclear reactions
are unstable and all the accumulated nuclear fuel is burned
explosively within a short time.  The result is a flash of thermal
X-ray emission from the surface of the star, which is called a Type I
X-ray burst.  The theory of these bursts has been developed by a
number of authors, e.g., Hansen \& van Horn (1975), Woosley \& Taam
(1976), Joss (1977), Fujimoto, Hanawa \& Miyaji (1981), Paczy\'nski
(1983), Fushiki \& Lamb (1987), Taam et al. (1993), Cumming \&
Bildsten (2000), Zingale et al.  (2001), Narayan \& Heyl (2003).  The
work described here is based on the method developed by Narayan \&
Heyl (2003), which has some advantages over the earlier work.

Figure 4 shows the results for the case of a $10M_\odot$ BHC with an
interior temperature of $10^{7.5}$ K (a reasonable choice, see Narayan
\& Heyl 2002).  We consider different choices for the radius $R_{\rm
BHC}$ of the black hole candidate, from $R_{\rm BHC}=(9/8)R_S$, the
smallest allowed value for a model in which there is no density
inversion with increasing radius (Weinberg 1972), up to $R_{\rm
BHC}\sim3R_S$, where $R_S$ is the Schwarzschild radius corresponding
to the mass of the BHC.  We have considered a range of mass accretion
rates and determined whether or not the system will have bursts.  The
solid dots in Fig. 4 indicate regions of the parameter space where the
system will produce bursts, and the empty regions indicate regions
where the accreting gas burns steadily without producing bursts.  It
is clear that in this model bursts are very common over a wide range
of accretion rates.

Figure 5 shows the expected recurrence times of bursts and the likely
durations of bursts (assuming the maximum luminosity during a burst is
equal to the Eddington luminosity) for a typical model with $R_{\rm
BHC}=2R_S$.  It is clear that for a range of accretion rates from a
tenth of Eddington upto almost the Eddington rate, bursts occur
reasonably frequently and have substantial fluences.  These bursts
would be hard to miss.  The fact that no bursts have been seen from
any BHC system in this accretion luminosity range is thus significant
and argues against such a model for BHCs (Narayan \& Heyl 2002).  For
accretion rates below a tenth Eddington, the systems are still
unstable (see Fig. 4), but the bursts are relatively rare (and have
correspondingly very large fluences).  The rarity of the bursts means
that this region of parameter space may be less useful for setting
constraints on models.

Another possibility for BHCs is the case we have focused on in this
paper, namely that the objects are fermion-fermion or boson-fermion
stars in which the gas collects as a compact sphere at the center of a
dark fermion or boson sphere.  Will such objects have Type I bursts
when they accrete?  To answer this question, we have redone the burst
calculations using the models described in \S\S2,3.  The quantities
required for the burst calculations are the ones shown in Figs. 2, 3,
namely the radius of the gas sphere $R_g$, the surface gravity $g$,
the redshift $z(R_g)$, and the accretion efficiency $\eta$.  With this
information, it is straightforward to take any given model of a
fermion-fermion or boson-fermion star, assume a local surface mass
accretion rate $\dot \Sigma$ (baryonic mass added per unit area per
unit time as measured in the local frame at the surface of the gas
sphere), and compute the thermonuclear stability of the accretion
layer.

The two top panels in Fig. 6 show the results for the fermion-fermion
star models described in \S2 and the two bottom panels correspond to
the boson-fermion models described in \S3.  In each panel, the
horizontal axis gives the accretion luminosity in Eddington units,
where $L_{\rm Edd}$ is defined at infinity for the total mass of the
star:
\begin{equation}
L_{\rm Edd} = {4\pi
G M_{\rmscr{grav}} c\over\kappa},
\end{equation}
with $\kappa =0.4 ~{\rm cm^2g^{-1}}$ (corresponding to electron
scattering in fully ionized hydrogen).  The vertical axis extends over
the full range of gas mass that we have considered in the hybrid
models.  The symbols (filled and open circles) in the panels
correspond to models that are predicted to exhibit bursts and the
empty spaces to models that are predicted not to have bursts.

We see that there is a considerable range of accretion rate for which
the systems will produce bursts.  Therefore, the absence of bursts in
BHCs argues against such a hybrid model for BHCs.  A comparison of
Figs. 4 and 6 shows that bursts occur in fermion-fermion and
boson-fermion models for lower accretion rates compared to BHC models
with hard surfaces.  This difference has an interesting explanation.
Recall that the Eddington luminosity was defined with respect to the
total mass of the hybrid star, which is the obvious choice for an
observer at infinity.  However, the accretion and burst physics occur
on the surface of the gas sphere which lies well within the dark
matter sphere.  Here, the local Eddington luminosity is given by
$L_{\rm Edd,gas}=4\pi R_g^2 cg/\kappa$, which is different from the
$L_{\rm Edd}$ defined earlier.  Redshifting this luminosity to the
observer at infinity gives
\begin{equation}
L_{\rm Edd,gas,\infty}={4\pi R_g^2 cg\over (1+z)^2\kappa},
\end{equation}
which is the physically correct luminosity to use as the scale when
analysing bursts from these objects.  We find that all the models we
have analysed are unstable for all accretion rates up to the local
Eddington limit.  Such models are shown by the solid circles in
Fig. 6.  In fact, the models are generally unstable even for slightly
super-Eddington local accretion rates, as shown by the open circles.

Figure 7 shows the predicted burst recurrence times and burst
durations for fermion-fermion and boson-fermion stars with a gas mass
of $0.3M_\odot$ (as an example).  The recurrence times are a little
long, but well within observational limits.  The burst durations have
been calculated with two different assumptions regarding the maximum
luminosity during the burst.  The solid circles assume that the
maximum luminosity is equal to the full Eddington luminosity $L_{\rm
Edd}$, while the open circles assume that the maximum luminosity is
equal only to $L_{\rm Edd,gas,\infty}$.  Which is the appropriate one
to use depends on whether or not the burst causes a substantial
``radius expansion'' of the radiating gas.  This cannot be determined
from the present calculations, but requires full time-dependent
computations of the development of the burst.

Regardless of the details, it appears that bursts should occur
reasonably frequently on fermion-fermion and boson-fermion stars and
that they should be easy to observe.  The absence of these bursts
means that such models may be ruled out for BHCs.

\section{Summary and Discussion}

For ordinary matter that is made up of baryons, and for any reasonable
equation of state, the maximum mass of a compact degenerate star is
about $3M_\odot$.  Black hole candidates (BHCs) in X-ray binaries
(XRBs) are more massive than this limit, so they cannot be baryonic.
However, we cannot rule out the possibility that they are made up of
some exotic non-baryonic form of matter.  Until we can eliminate this
possibility, we cannot be certain that BHCs are true black holes.  We
have considered two kinds of exotic stars in this paper.

In one kind of model, we assume that the exotic star has a hard
surface on which normal accreting gas can pile up.  Such an object
should behave very much like a neutron star except that it will have a
different radius and surface gravity.  Narayan \& Heyl (2002) showed
that such objects would exhibit Type I X-ray bursts when they accrete
gas from a companion star.  We have here repeated the calculations
using the improved version of the burst calculations described in
Narayan \& Heyl (2003).  The main freedom in the models is the radius
that we assume for the surface of the BHC.  We have tried a wide range
(see Fig. 4) and find that the results are not sensitive to the choice
of radius.  The calculations indicate that $10M_\odot$ BHCs with hard
surfaces are unstable to bursts over an even wider range of mass
accretion rates than neutron stars (compare Fig. 4 in this paper to
Fig. 10 in Narayan \& Heyl 2003).  The burst durations and recurrence
times for the BHCs are longer by a factor of a few than the
corresponding time scales for neutron stars, but they are well within
the reach of observations.

An alternative variety of exotic star involves dark fermion or boson
matter that does not interact with gas except via gravity.  The lack
of interaction means that accreting gas sinks to the center and forms
a separate fluid component at the center of the dark matter
distribution.  The main free parameter in these models is the mass of
the fermion or boson particles in the dark matter.  There is not too
much freedom, however, because we require the dark matter sphere to
have a radius of no more than several Schwarzschild radii.  This limit
comes from the fact that observations of the X-ray emission of BH XRBs
constrain the radius of the accretion disk to be quite small, no
larger than about $6GM/c^2$ (McClintock \& Remillard 2003).  High
frequency quasi-periodic oscillations, in particular, require that
most of the mass of the compact star should be inside this radius,
which strongly constrains possible models of BHCs.  In view of this
constraint, we find that viable models are possible only with fermions
of mass $\sim200$ MeV or bosons of mass $\sim {\rm few}\times10^{-17}$
MeV.  While we have done the calculations for one particular choice of
the mass in each case (223 MeV and $2.4\times10^{-17}$ MeV,
respectively), we expect the results to be similar for other choices
within the relevant range.

We have considered model objects consisting of $10M_\odot$ of dark
matter (fermions or bosons) and a wide range of mass in the gas
component: $10^{-6}M_\odot$ up to about $1M_\odot$, the maximum mass
beyond which the models become black holes.  Over this entire range of
models, we have calculated the thermonuclear stability of gas
accreting on the surface of the gas sphere.  In all cases, we find
that Type I bursts are expected.  Indeed, these models burst for all
acretion rates up to the local Eddington rate as calculated at the
surface of the gas sphere.  (This is less than the Eddington rate for
the combined gas plus dark matter object, see \S4.)  Moreover, the
burst durations and recurrence times are not very unusual (Fig. 7
shows results for the specific case when the gas mass is
$0.3M_\odot$), and the bursts should be easy to observe.

What is a reasonable mass for the gas sphere in the models?  The
answer of course depends on how the particular exotic star is formed.
If the star results from the death of a normal star, in which much of
the gas is converted somehow to exotic matter, then one imagines that
a reasonable fraction of the gas (perhaps several tenths of a solar
mass?) might survive as normal matter.  In this case, the upper range
of gas masses that we have considered would be appropriate.  On the
other hand, if the exotic star is somehow born as a pristine purely
dark matter object, then the only gas it would contain is whatever it
accumulates through accretion.  The typical mass transfer rate from
the companion star in black hole XRBs is $\dot M_{\rm transfer} \sim
10^{-10} - 10^{-9} ~M_\odot {\rm yr^{-1}}$ (King, Kolb \& Burderi
1996), and the typical lifetime of the systems is $\sim 10^8 - 10^9$
yr.  Thus, we expect the accreted mass to be $\sim0.1M_\odot$.  This
is again near the upper end of the range we have considered (we went
as low as $10^{-6}M_\odot$, see Fig. 6).

The basic result of our study is that, for any kind of exotic star
that we can consider, Type I X-ray bursts are expected to be present.
Can these bursts be seen, and if so why have they not been observed?
Tomsick et al. (2003) recently reported detecting bright $\sim 100$ s
flares in the X-ray flux of the BH XRB XTE J1650-500.  These
particular flares had nonthermal energy spectra, so they were not Type
I X-ray bursts (which always have thermal spectra as one would expect
for emission from a stellar surface).  However, this work demonstrates
that bursts are easy to detect in BHCs.  The fact that no bona fide
Type I burst has been reported in any of the dozen or more BHCs is
thus highly significant.  It would be useful to go back to the
archival data on these systems and to derive quantitative limits on
the burst rate.

To our knowledge, only one study so far --- Tournear et al. (2003) ---
has specifically looked for bursts in BHCs.  This work was done using
data collected with the {\it Unconventional Stellar Aspect Experiment}
and the {\it Rosse X-Ray Timing Explorer}.  The authors derive a 95\%
upper limit on the burst rate in BHCs of $2.0\times10^{-6} ~{\rm
bursts \,s^{-1}}$, compared to a mean detection rate in neutron star
systems of $(4.1\pm0.8)\times10^{-5} ~{\rm bursts\,s^{-1}}$, which is
about 20 times higher.  Since our calculations indicate that the burst
rate in BHCs (if the objects have surfaces) should be only a factor of
a few lower than in neutron stars, the upper limit that Tournear et
al.  have obtained is already quite interesting.  A longer observation
might lead to quite strict limits on bursts in BHCs.

The most obvious explanation for the lack of bursts in BHCs is that
the objects are true black holes, with event horizons.  Narayan (2003)
has discussed various other explanations for the lack of bursts in
BHCs and argues against all except the following two.  One possibility
is that BHCs are exotic stars with non-interacting dark matter through
which gas can percolate down to the center.  This is the model that we
have analysed in the present paper.  We have shown that such a model
will produce observable bursts and is therefore not viable.  The other
possibility discussed by Narayan (2003) is that the unstable accretion
layer on the surface of a BHC does not burst coherently in large
explosions, in which the entire surface participates, but rather goes
off in a series of random localized mini-explosions.  The latter would
be hard to distinguish observationally.  Mini-explosions are expected
if the burning front is unable to propagate rapidly over the surface
of the star.  It is hard to eliminate this possibility since the
physics of deflagration fronts is not fully understood (but see
Spitkowsky, Levin \& Ushomirsky 2002).  However, for a given nuclear
fuel, we expect the propagation speed of the front to depend primarily
on the local surface gravitational acceleration.  Since $g$ on the
surface of a BHC is somewhat below that on a neutron star, and since
both neutron stars and white dwarfs (which have much smaller values of
$g$) are able to burst coherently, it is hard to see why BHCs should
have any particular difficulty in producing large coherent bursts.

Another caveat worth mentioning is that, in Fig. 4, we did not
consider radii for the BHC smaller than $(9/8)R_S$.  This radius is
the smallest allowed within General Relativity for an object with a
physical density distribution that satisfies $d\rho/dr \leq 0$
(Weinberg 1972).  If we allow a positive density gradient --- which
risks the Rayleigh-Taylor instability and requires a non-monotonic
behavior of pressure with density --- then smaller radii are possible.
For a sufficiently small radius, the redshift would become very large
and it would be possible to hide bursts from the view of the observer.
As Abramowicz, Kluzniak \& Lasota (2002) have noted, the gravitational
condensate star described by Mazur \& Mottola (2002) is one such model
that can have an extremely large redshift.  However, this particular
model has serious conceptual problems.

Finally, we have assumed that the gas is fully degenerate, whereas one
might wonder whether the gas component could remain hot and extended,
and be thermally supported, for the entire age of the system (like
low-mass brown dwarfs which retain their internal energy for nearly a
Hubble time).  Fortunately, this kind of a model may be ruled out.  As
mentioned earlier, X-ray observations indicate a small radius for the
accretion disk, which means that the gas sphere must have an even
smaller radius.  For such radii, the virial temperature is very high
($\sim10^{11}$ K), and the density of the gas is also large.  Neutrino
cooling is then highly effective and the gas loses most of its thermal
energy in a very short time (see Shapiro \& Teukolsky 1983 for a
discussion of the related problem of neutron star cooling).  The gas
thus quickly achieves degenerate conditions, where our calculations
apply.

\acknowledgements The authors thank Jeremy Heyl for useful
discussions.  This work was supported in part by NASA grant NAG
5-10780 and NSF grant AST 0307433. Y.F.Y is partially supported 
by the Special Funds for Major State Research Projects, and the 
National Natural Science Foundation (10233030).

\bibliographystyle{apj}
\bibliography{ns,mine,lmxb,typei,fus,math}

\begin{appendix}
\section{Accretion Efficiency of a Fermion-Fermion Star}

We begin by quoting a result from Rosenbluth et al. (1973).  Consider
a gravitating star that is made of a single fluid whose internal
energy $e$ is a function only of the density $\rho$ and satisfies $de
= -pd(1/\rho)$.
If $E(M)$ is the binding energy for a given mass $M$,
Rosenbluth et al. (1973) show via a Newtonian analysis that 
\be
{d E\over d M} = \Phi_s, 
\ee 
where $\Phi_s$ is the gravitational potential at the surface of the
star.  Thus, if a small quantity of mass $\Delta M$ is added to the
star, the change in energy is equal to $\Phi_s\Delta M$.  Since this
quantity is equal to the energy released in dropping the mass from
infinity down to the surface, equation (A1) implies that there is no
additional internal energy released when the star adjusts to its new
equilibrium.  In a sense, the result is not surprising.  Stars of
different $M$ all have the same specific entropy (because of the
particular form of the equation of state), so there is neither an excess
nor a deficit of specific entropy in going from one stellar mass $M$
to another.

We are interested in generalizing the above result for the
relativistic problem, and also for the case of two fluids.  Let us
first consider a general relativistic star with a single fluid.  Let
$M_{\rm grav}$ and $M_{\rm bar}$ be the gravitational and baryonic masses of
the star and let $R$ be its radius.  Let the fluid consist of
particles of mass $m$ with a number density $n(r)$.  Write the metric
function $A(r)$ as 
\be 
A(r) \equiv \left[1-{2M(r)\over r}\right]^{-1}, \qquad 
A(R) = \left(1-{2M_{\rmscr{grav}}\over R} \right)^{-1}, 
\ee
where we have used gravitational units with $G=c=1$.  The
gravitational and baryonic masses of the star are given by
\be
M_{\rmscr{grav}} = \int_0^R 4\pi r^2 \rho dr, \ee \be M_{\rmscr{bar}} = \int_0^R 4\pi
r^2 mn (1-2M/r)^{-1/2} dr.
\ee
In equilibrium, for
a given baryonic mass, the star will take up the configuration with
the lowest energy, i.e., the smallest value of $M_{\rm grav}$.  Thus, for
small variations of the equilibrium, we must have
\be
\delta M_{\rmscr{grav}} - \lambda \delta M_{\rmscr{bar}} = 0,
\ee
where $\lambda=\partial M_{\rm grav}/\partial M_{\rm bar}$ is a Lagrange
multiplier.  

Equation (A5) must be valid for all first-order
variations.  Consider a specific variation consisting of a Dirac
delta-function at the surface of the star,
\be
\delta n(r) = \epsilon \delta (r-R), \qquad
\delta \rho(r) = m\epsilon \delta (r-R),
\ee
where the second relation follows from the fact that $n$ and $\rho$
both go to zero at the surface, so that the fluid has no internal
energy in this limit.  Substituting (A6) into (A3) and (A4), we find
\be
\delta M_{\rmscr{grav}} = \epsilon m 4\pi R^2, \qquad
\delta M_{\rmscr{bar}} = \epsilon m 4\pi R^2 (1-2M_{\rmscr{grav}}/R)^{-1/2}.
\ee
Substituting this in equation (A5), we then find
\be
\lambda = {\partial M_{\rmscr{grav}}\over \partial M_{\rmscr{bar}}} =
\left(1-{2M_{\rmscr{grav}}\over R}\right)^{1/2} = {1\over 1+z(R)},
\ee
where $z(R)$ is the gravitational redshift at the surface of the star.
This result is an obvious and natural generalization of the Newtonian
result (A1).

Consider now the problem of interest, a two fluid fermion-fermion
star.  Let us use subscripts 1 and 2 for the two fluids and assume
that the outer radius $R_1$ of fluid 1 is smaller than the radius $R_2$
of fluid 2.  (In the context of the main paper, fluid 1 is the gas and
fluid 2 is the dark matter.)  Define the gravitational mass
at a general radius $r$ as
\be
M(r) = \int_0^r 4\pi r'^2(\rho_1+\rho_2)dr',
\ee
so that the total gravitational mass of the star is
\be
M_{\rmscr{grav}} = M(R_2).
\ee
The baryonic masses of the two fluids are given by
\be
M_{\rmscr{bar,1}} = \int_0^{R_1} 4\pi r^2m_1n_1(1-2M/r)^{-1/2}dr,
\ee
\be
M_{\rmscr{bar,2}} = \int_0^{R_2} 4\pi r^2m_2n_2(1-2M/r)^{-1/2}dr.
\ee
Requiring the total gravitational mass $M_{\rm grav}$ to be a minimum
for fixed baryonic masses of the two
fluids gives the variational condition,
\be
\delta M_{\rmscr{grav}} - \lambda_1\delta M_{\rmscr{bar,1}} - \lambda_2 \delta
M_{\rmscr{bar,2}} = 0,
\ee
where $\lambda_1$ and $\lambda_2$ are two Lagrange multipliers.

Consider first a perturbation of the form
\be
\delta n_1(r) = 0, \qquad \delta n_2(r) = \epsilon \delta (r-R_2).
\ee
Substituting this in the above equations, and following the same steps
as for the single fluid case, we directly obtain
\be
\lambda_2 = \left(1-{2M_{\rmscr{grav}}\over R_2}\right)^{1/2} 
= {1\over 1+z(R_2)},
\ee
which is identical to the result for a single fluid star.

Consider next a perturbation
\be
\delta n_1(r) = \epsilon\delta(r-R_1), \qquad \delta n_2(r) = 0.
\ee
This gives
\be
\delta M(r) = \epsilon m_1 4\pi R_1^2, \quad r\geq R_1,
\ee
\be
\delta M_{\rmscr{grav}} = \epsilon m_1 4\pi R_1^2,
\ee
\be
\delta M_{\rmscr{bar,1}} = \epsilon m_1 4\pi R_1^2(1-2M/R_1)^{-1/2},
\ee
\be
\delta M_{\rmscr{bar,2}} = \int_{R_1}^{R_2} 4\pi rm_2n_2(1-2M/r)^{-3/2}
\epsilon m_1 4\pi R_1^2 dr.
\ee
Substituting these relations in equation (A13) and making use of 
equation (A15), we obtain the following result for
$\lambda_1$:
\be
\lambda_1 = \left[ 1-{2M(R_1)\over R_1} \right]^{1/2} \left[1-\left(
1-{2M_{\rmscr{grav}}\over R_2} \right)^{1/2}
\int_{R_1}^{R_2}4\pi rm_2n_2\left( 1-{2M\over r}\right)^{-3/2}dr \right].
\label{eq:lambda1}
\ee
This expression does not look as simple as equations (A8) or (A15),
but in fact it simplifies considerably when written in terms of the
redshift.

To see this, start with the well-known Tolman-Oppenheimer-Volkoff
structure equations  for a relativisic star.  Restricting ourselves
to radii $r\geq R_1$, where there is only fluid 2, the equations take 
the form
\bea
\frac{dp_2}{dr}&=&-\frac{(\rho_2+p_2)(M+4\pi r^3 p_2)}{r^2(1-2M/r)}, \label{eq:TOV}\\
\frac{d \ln \sqrt{B}}{dr}&=&\frac{1}{\rho_2+p_2}\frac{dp_2}{dr}. \label{eq:dBdr}
\eea
Using the thermodynamic relation,
\be
d\rho_2=\frac{\rho_2+p_2}{n_2}dn_2, \label{eq:eos2}
\ee
the above equations can be rewritten as
\be
\frac{d}{dr}\left[\frac{n_2}{\rho_2+p_2}\left(1-\frac{2M}{r}\right)^{-1/2} 
\right] 
	= 4\pi r n_2 \left(1-\frac{2M}{r}\right)^{-3/2} ,\label{eq:TOV2}
\ee
\be
\frac{d \ln \sqrt{B}}{dr} = \frac{d}{dr}\ln \left(\frac{n_2}{\rho_2+p_2} \right). \label{eq:dBdr2}
\ee
Integrating these equations and evaluating the result at $r=R_1$, we find
\bea
\frac{n_2(R_1)m_2}{\rho_2(R_1)+p_2(R_1)}&=&\left(1-\frac{2M(R_1)}{R_1} \right)^{1/2}
        \left(1-\frac{2M_{\rmscr{grav}}}{R_2} \right)^{-1/2} \nonumber \\
&&      \left[1-\left(1-\frac{2M_{\rmscr{grav}}}{R_2}
\right)^{1/2} \int_{R_1}^{R_2} 4 \pi r m_2 n_2 \left(1-\frac{2M}{r} \right)^{-3/2}
dr\right] \label{eq:TOV3} ,\\
\sqrt{B(R_1)}&=&\left(1-\frac{2M(R_2)}{R_2} \right)^{1/2}\frac{n_2(R_1)m_2}{\rho_2(R_1)+p_2(R_1)}. \label{eq:dBdr3}
\eea

Finally, substituting (A27) in (A28) and comparing with (A21), we find
that 
\be
\lambda_1 = \frac{\partial M_{\rmscr{grav}}}{\partial M_{\rmscr{bar,1}}} 
= \sqrt{B(R_1)} = {1\over
1+z(R_1)}, 
\ee 
i.e., when written in terms of the redshift, the formula for
$\lambda_1$ for the two-fluid case is identical to that for a 
single fluid.  The accretion efficiency $\eta$ defined in the text
is then given by
\be 
\eta = 1-\frac{\partial M_{\rmscr{grav}}}{\partial M_{\rmscr{bar,1}}} 
= {z(R_1)\over 1+z(R_1)}.
\ee 

We have confirmed equation (A29) by numerically computing the partial
derivative $\partial M_{\rm grav}/\partial M_{\rm bar,1}$ and comparing
it to $1/[1+z(R_1)]$.  The agreement is very good.  Indeed, numerical
calculations show that equation (A29) is true also for a boson-fermion
star, though we do not have a formal analytical proof for that case.

\end{appendix}

\begin{figure}
\epsscale{0.8}
\plotone{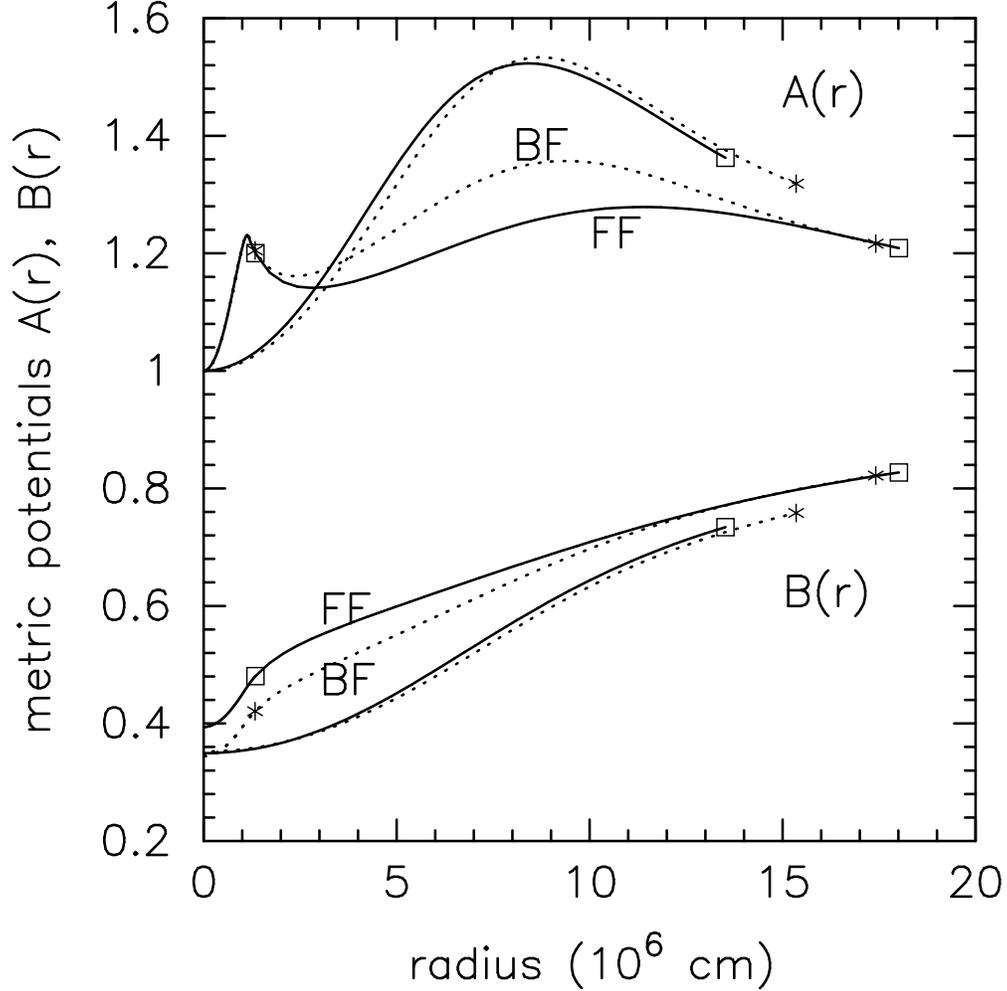}
\caption{ Variation of the metric potentials $A(r)$ and $B(r)$ with
radius for representative models.  Solid and dotted lines show the
results for fermion stars and boson stars, respectively.  The curves
identified as ``FF'' show the results for fermion-fermion stars with
$10M_\odot$ of fermionic dark matter and $0.7M_\odot$ of normal gas,
while the curves identified as ``BF'' show the results for
boson-fermion stars with the same component masses.  The curves not
identified as FF or BF correspond to the pure dark matter stars with
no gas.  The squares show the surfaces of the gas and fermion
components in the fermion stars, and the stars show the surfaces of
the gas and boson components in the boson stars.  For the boson
component, the surface is defined as the radius that encloses 99.9\%
of the boson mass.}
\label{fig:metric}
\end{figure}

\begin{figure}
\plotone{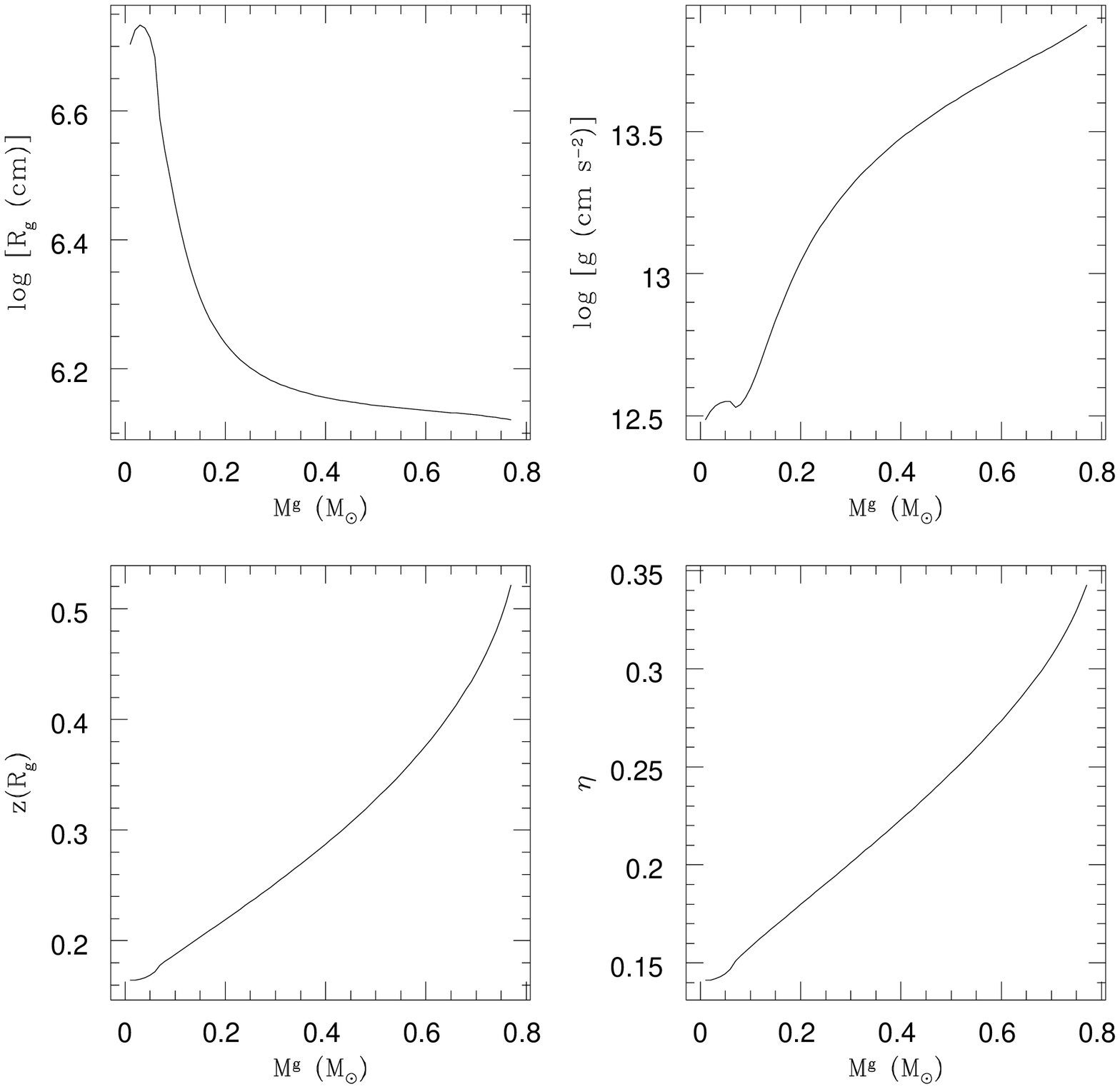}
\caption{Properties of the gas component in fermion-fermion stars.
The four panels show the variations of the radius of the gas sphere
$R_g$, the surface gravitational acceleration $g$, the surface
redshift $z(R_g)$, and the dimensionless binding energy at the surface
$\eta$, as functions of the baryonic gas mass $M^g$.  The fermionic
dark matter component has a fixed mass of $10M_\odot$ in the models.}
\label{fig:fermionhybrid}
\end{figure}

\begin{figure}
\plotone{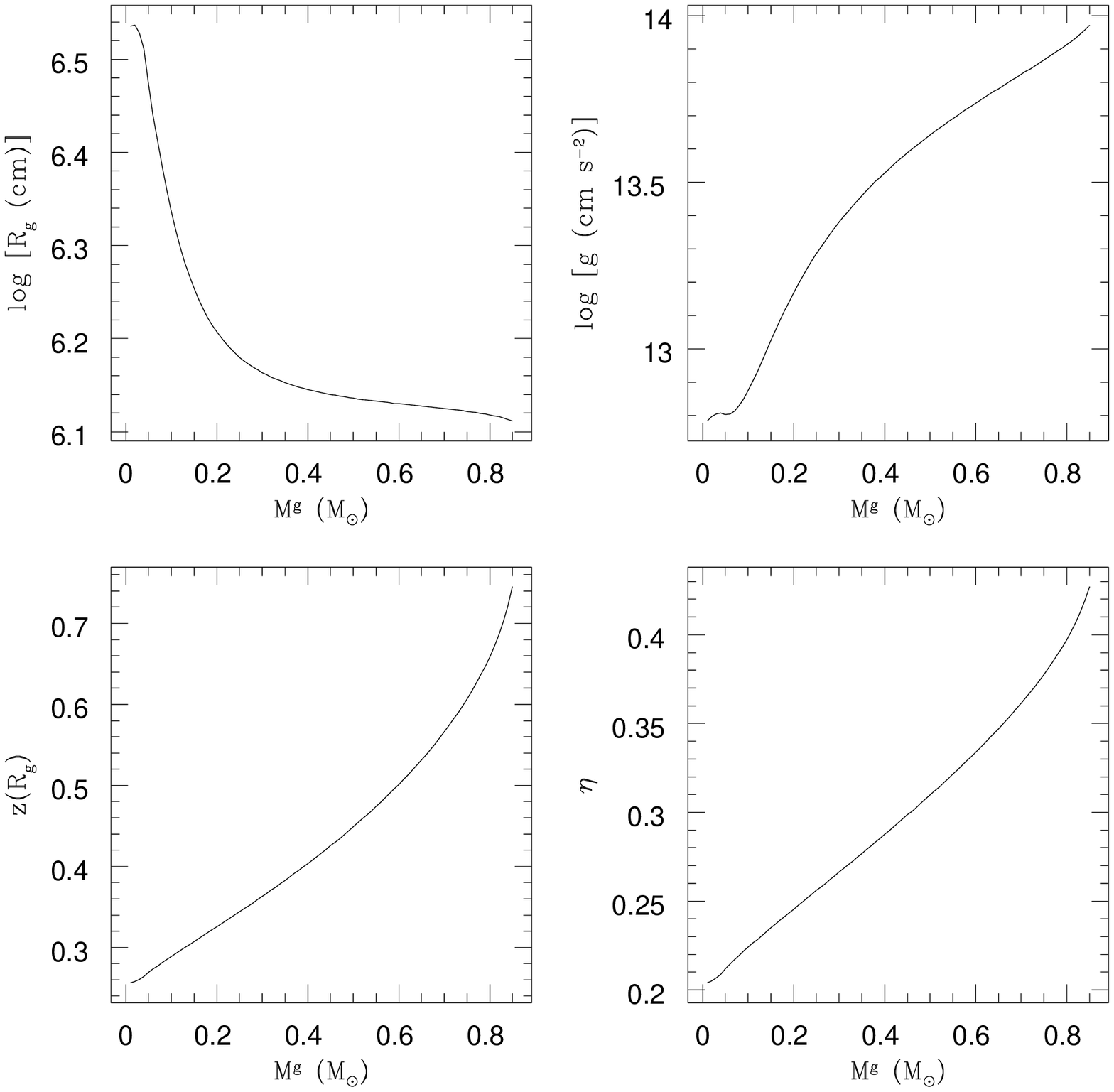}
\caption{Properties of the gas component in boson-fermion stars.  The
four panels show the variations of the radius of the gas sphere $R_g$,
the surface gravitational acceleration $g$, the surface redshift
$z(R_g)$, and the dimensionless binding energy at the surface $\eta$,
as functions of the baryonic gas mass $M^g$.  The bosonic dark matter
component has a fixed mass of $10M_\odot$ in the models.}
\label{fig:Bosonhybrid}
\end{figure}

\begin{figure}
\plotone{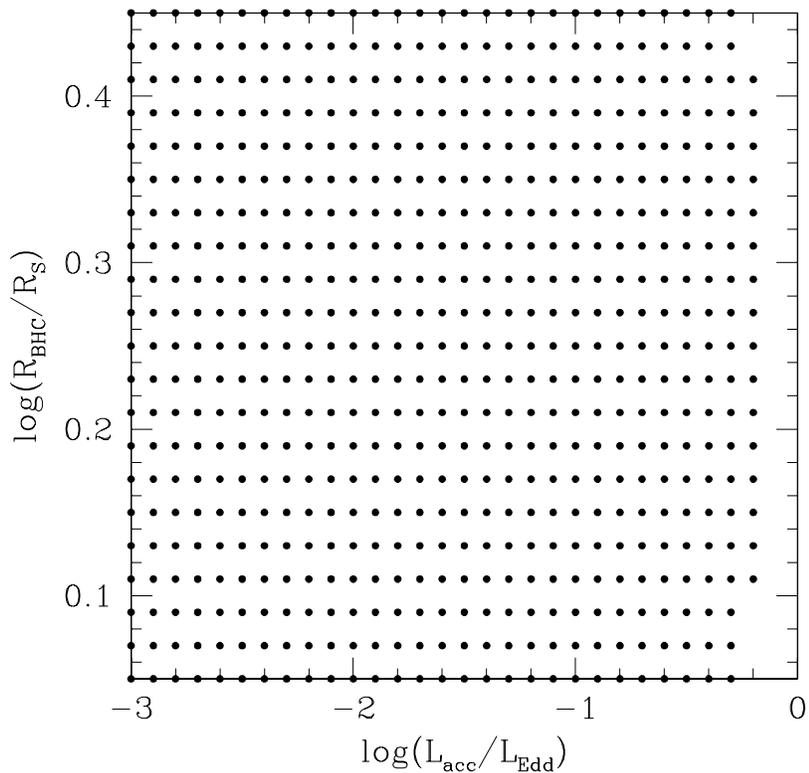}
\caption{Burst stability of a $10M_\odot$ black hole candidate with a
hard surface.  The accretion rate in Eddington units is shown along
the abscissa and the radius of the object in Schwarzschild units is
shown along the ordinate.  The dots represent models that are unstable
to bursts, and the empty region (very close to the Eddington limit)
corresponds to models that are stable and do not burst.  Note that
there is very little parameter space where bursts are absent.}
\label{fig:BHCsurface}
\end{figure}

\begin{figure}
\plotone{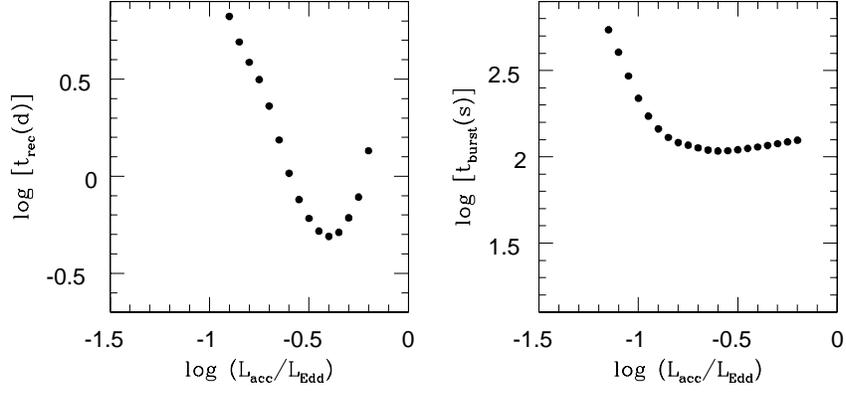}
\caption{Left: Shows the variation of the burst recurrence time
$t_{\rm rec}$ in days as a function of the accretion rate for a
$10M_\odot$ object with a hard surface.  A radius of $2R_S = 59$ km
has been assumed.  Bursts are readily observable for accretion rates
in the range from about $0.1L_{\rm Edd}$ to about $0.7L_{\rm Edd}$.
Right: Shows the variation of the burst duration $t_{\rm burst}$ in
seconds for the same models.  The durations have been computed
assuming that the fluence in the burst emerges at the Eddington
luminosity.}
\label{fig:BHCsurface2}
\end{figure}

\begin{figure}
\plotone{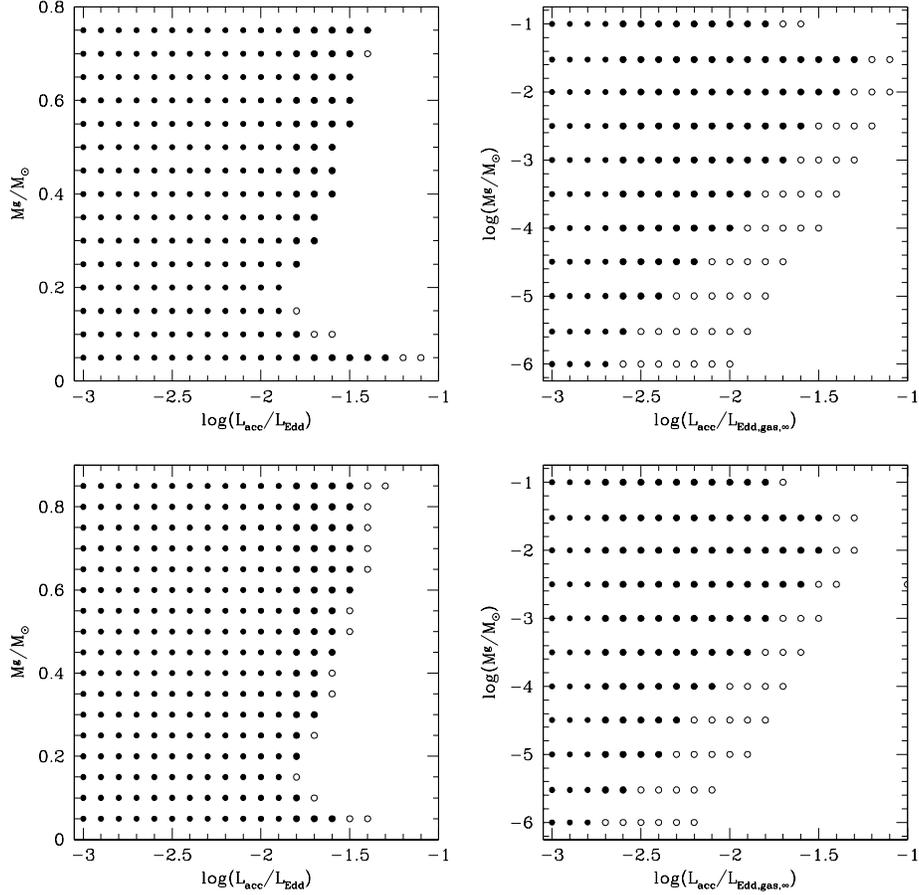}
\caption{Top Left: Burst stability for fermion-fermion stars as a
function of the accretion rate in Eddington units along the abscissa
(calculated for the total mass of the object) and the mass of the gas
component along the ordinate.  The symbols correspond to models that
are unstable to bursts, and the empty regions to models that are
stable to bursts.  Filled circles represent models that accrete at
less than the local Eddington rate at the surface of the gas
component.  Open circles represent models that accrete at above the
local Eddington rate but still well below the Eddington rate for the
total mass.  Top Right: Similar to Top Left, but for small gas masses
(on a logarithmic scale) in the range $10^{-6}-10^{-1}M_\odot$.
Bottom Left, Bottom Right: Corresponding results for boson-fermion
stars.}
\label{fig:FermionBoson}
\end{figure}

\begin{figure}
\plotone{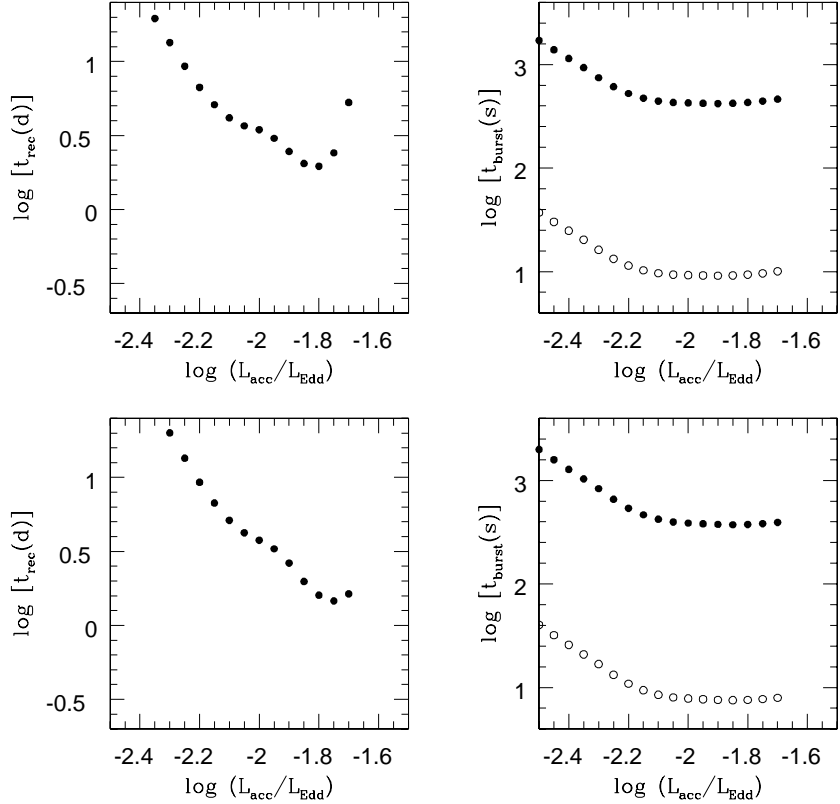}
\caption{Top Left: Burst recurrence time $t_{\rm rec}$ in days as a
function of the accretion rate for a fermion-fermion star with a gas
mass of $M^g=0.3M_\odot$.  Top Right: Burst duration $t_{\rm burst}$
for the same models.  The durations have been computed by assuming
that the fluence in the burst is emitted at the Eddington luminosity;
the filled circles are the results when the local Eddington luminosity
at the surface of the gas sphere is used and the open circles when the
Eddington luminosity of the total mass is used.  Bottom Left, Bottom
Right: Corresponding results for a boson-fermion star of the same
mass.}
\label{fig:FermionBoson2}
\end{figure}

\end{document}